\documentclass[fleqn,usenatbib]{mnras}

\usepackage[T1]{fontenc}
\usepackage{ae,aecompl}

\usepackage{savesym}  
\usepackage{graphicx}	
\usepackage{amsmath}	
\usepackage{amssymb}	
\usepackage{natbib}
\savesymbol{iint}  
\usepackage{txfonts}  
\restoresymbol{TXF}{iint}  

\usepackage[british]{babel}             

\newcommand{\cii}{[C\,{\sc ii}] }

\title[ {\it [C\,{\sc ii}]} emission in $z\sim 6$ strongly lensed galaxies]{[C\,{\sc ii}] emission in $z\sim 6$ strongly lensed, star-forming galaxies}

\author[K.K. Knudsen et al.]{Kirsten K. Knudsen,$^{1}$\thanks{E-mail: kirsten.knudsen@chalmers.se}
Johan Richard,$^{2}$ 
Jean-Paul Kneib,$^{3,4}$
Mathilde Jauzac,$^{5,6,7}$ 
\newauthor
Benjamin Cl\'ement,$^{2}$
Guillaume Drouart,$^{8}$ 
Eiichi Egami,$^{9}$
Lukas Lindroos$^{1}$
\\
$^{1}$Department of Earth and Space Sciences, Chalmers University
of Technology, Onsala Space Observatory, SE-43992 Onsala, Sweden \\
$^{2}$Univ Lyon, Univ Lyon1, Ens de Lyon, CNRS, Centre de Recherche
Astrophysique de Lyon UMR5574, F-69230, Saint-Genis-Laval, France \\
$^{3}$Laboratoire d'Astrophysique, Ecole Polytechnique F\'ed\'orale de
Lausanne (EPFL), Observatoire de Sauverny, CH-1290 Versoix, Switzerland \\
$^{4}$Aix Marseille Universit\'e , CNRS, LAM (Laboratoire d'Astrophysique de
Marseille), UMR 7326, 13388 Marseille, France\\
$^{5}$Centre for Extragalactic Astronomy, Department of Physics, Durham
University, Durham DH1 3LE, U.K.\\
$^{6}$Institute for Computational Cosmology, Durham University, South Road,
Durham DH1 3LE, U.K.\\
$^{7}$Astrophysics and Cosmology Research Unit, School of Mathematical
Sciences, University of KwaZulu-Natal, Durban 4041, South Africa\\
$^{8}$International Centre for Radio Astronomy Research, Curtin University,
Bentley, WA6102, Perth, Australia \\
$^{9}$Steward Observatory, University of Arizona, 933 N. Cherry Ave, Tucson,
AZ 85721, USA 
}

\date{Accepted XXX. Received YYY; in original form ZZZ}

\pubyear{2016}

\begin{document}
\label{firstpage}
\pagerange{\pageref{firstpage}--\pageref{lastpage}}
\maketitle

\begin{abstract}
The far-infrared fine-structure line \cii at 1900.5\,GHz is known 
to be one of the brightest cooling lines in local galaxies, and therefore it
has been
suggested to be an efficient tracer for star-formation in very high-redshift
galaxies.  However, recent results for galaxies at $z>6$ have yielded numerous
non-detections in star-forming galaxies, except for quasars and submillimeter
galaxies. 
We report the results of ALMA observations of two lensed, star-forming
galaxies at $z = 6.029$ and $z=6.703$.  The galaxy A383-5.1 (star formation
rate [SFR] of 3.2\,M$_\odot$\,yr$^{-1}$ and magnification of $\mu =
11.4\pm1.9$) 
shows a line
detection with $L_{\rm [CII]} = 8.9\times10^{6}$\,L$_\odot$, making it the
lowest $L_{\rm [CII]}$ detection at $z>6$.  For MS0451-H 
(SFR\,=\,0.4\,M$_\odot$\,yr$^{-1}$ and $\mu = 100\pm20$) we provide an upper limit of
$L_{\rm [CII]} < 3\times10^{5}$\,L$_\odot$,
which is 1\,dex below the local SFR-$L_{\rm [CII]}$ relations.  The results
are consistent with predictions for low-metallicity galaxies at $z>6$,
however, other effects could also play a role in terms of decreasing $L_{\rm
[CII]}$.  The detection of A383-5.1 is encouraging and suggests that
detections are possible, but much fainter than initially predicted.   
\end{abstract}



%

\section{Introduction}

During the past decade, the number of galaxies with measured redshifts $z> 6$
has increased significantly \citep[e.g.][]{hu02,iye06} 
with even a few spectroscopic redshifts of $z>7$
\citep{vanzella11,ono12,schenker12,shibuya12,finkelstein13,watson15,oesch15,zitrin15}, 
demonstrating the tremendous potential for progress of our understanding of 
galaxy formation during the first billion years after the Big Bang. 
There is an even larger number of galaxies with photometric redshifts $z>6$
\citep[e.g.][]{mclure13,smit15},
however, spectroscopic redshifts are essential for studying the
physical properties of the interstellar medium and the gas that fuels the star
formation.  
Because of the intrinsic high luminosity and the large gas masses, several
starburst and quasar host galaxies have been studied in great detail at $z>6$
\citep[e.g.][]{maiolino05,venemans12,wang13,riechers13,willott15a,banados15,cicone15,venemans15}.
Such galaxies are interesting to understand the evolution of the
most massive galaxies but are not representative of the overall galaxy
population. 

Various tracers are used for determining the properties of the gas 
in star-forming galaxies.  In terms of studying molecular gas, CO is most
commonly used as it is the second most abundant molecule and with bright
emission lines from the rotational transitions that are visible in the
(sub-)mm bands.  
One of the brightest lines seen in the far-infrared (FIR) in local
star-forming galaxies is the fine-structure (FS) line \cii ($^2 P _{3/2} \to ^2
P_{1/2}$) at 1900.537\,GHz, which is found to correlate
with the star formation rate \citep[e.g.][]{delooze14,sargsyan14}.  While
local studies of the FIR FS lines are difficult due to
the opacity of the Earth's atmosphere, 
at high $z$ the lines are shifted towards the (sub-)mm bands and thus
observable with ground-based telescopes.  

With the increased ground-based capabilities of mm-wavelength telescopes,
it has become possible to search for the \cii line in less
extreme galaxies at $z>6$. However, a clear picture of this line as a 
tracer of the star formation is not emerging.  
Recently, \citet{capak15} and \citet{willott15b} 
detected \cii in bright Lyman-break galaxies (LBG) at $5<z<6$
and $z\sim6.1$, respectively, with the \cii line
luminosities and SFR following similar relations as those found for local
star-forming galaxies. The galaxies are UV-luminous (1-4$L^\ast$; we use
$L^\ast$ for the corresponding redshift),
representing the bright end of the UV-luminosity function.  
\citet{maiolino15} targeted three LBGs within the redshift range $z=6.8-7.1$,
but did not detect the \cii line despite the galaxies having estimated SFRs
$\sim 5-15$\,M$_\odot$\,yr$^{-1}$.  However, they find a detection that is 
offset from one of the targets possibly explained by feedback and/or gas
accretion.  Similarly, \citet{schaerer15} obtained upper limits for two other
LBGs at $z\sim 6.5-7.5$, of which one is lensed and with sub-$L^\ast$
luminosity.  
Ly$\alpha$ emitting galaxies have been observed,
including the massive Ly$\alpha$-blob 'Himiko', however, no confirmed
detection has so far been obtained
\citep[e.g.][]{ouchi13,kanekar13,ota14,gonzalez14}.  
Almost all of these galaxies have non-detections in the FIR 
continuum suggesting a low dust-mass.  \citet{watson15}
found a clear detection of dust emission from a spectroscopically
confirmed $z=7.5$ galaxy, however, did not detect an expected bright \cii
line in the frequency range covered.  The galaxy is strongly lensed by a
magnification factor of 9.5, thus providing constraints on a sub-$L^\ast$ galaxy.  

In this letter we present ALMA observations of \cii for two
sub-$L^\ast$ galaxies at $z>6$.  The two sources are A383-5.1 ($z = 6.027$) 
and MS0451-H ($z=6.703$),
which have estimated SFRs of 3.2 and 0.4\,M$_\odot$\,yr$^{-1}$, respectively,
and estimated magnification factors of $\mu = 11.4\pm1.9$ and $100\pm20$,
respectively \citep[][Kneib et al. in preparation]{richard11,stark15}.  
The strong lensing of both galaxies enables us to probe intrinsically fainter
luminosities and SFR than previous
observations.  
We assume a $\Lambda$CDM cosmology with $H_0 =
67.3$\,km\,s$^{-1}$\,Mpc$^{-1}$, $\Omega_M = 0.315$, and $\Omega_\Lambda =
0.685$ \citep{planck13}.  

%

\section{Observations}


We observed the sources MS0451$-$H and A383$-$5.1 with ALMA in Cycle-2. 
The observations were carried out in December 2014, January and May
2015.
A separate receiver setup was used for each source tuning to the redshifted
\cii line; for A383 at $z = 6.027$ \footnote{The setup for A383-5.1 was based
on the redshift  from \citet{richard11}, the redshift was
revised to 6.029 following observations of the counter-image A383-5.2
\citep{stark15}.} the central frequency was 270.462\,GHz
and for MS0451-H at $z = 6.703$ it was 246.727\,GHz.  
The correlator was used in the frequency domain mode with one spectral window
(spw)
having a bandwidth of 1.875\,GHz centered on the aforementioned frequencies,
and the three other available spw's used a continuum setup with a bandwidth of 2\,GHz each
distributed over 128 channels. 
The telescope configuration has baselines extending between 15 and 350\,m for
MS0451-H and 15 to 540\,m for A383-5.1; 
the observations include baselines longer than initially proposed for.   
The integration time was 1.9 hours for MS0451-H and 4.6 hours for A383-5.1. 
Table \ref{tab:obssum} summarizes the details of the observations including
a list of the calibrators.  

Reduction, calibration, and imaging was done using {\sc casa} (Common
Astronomy Software Application\footnote{https://casa.nrao.edu};
\citealt{mcmullin07}).  
The pipeline reduced data delivered from the observatory was of sufficient
quality, no additional flagging and further calibration was necessary.  
The pipeline includes the steps required for standard reduction and
calibration, such as flagging, bandpass calibration, as well
as flux and gain calibration.  
A conservative estimate of the absolute flux calibration is $10\%$. 
In the case of A383-5.1, a continuum source, flux density $\sim 2$\,mJy, is
seen 11\,arcsec south of the point center.  We attempted to self-calibrate
using this relatively bright source, however, this did not significantly
improve the sensitivity.  

The data was imaged both as continuum and spectral cube using natural
weighting.  A continuum image was produced combining all spectral windows,
while a spectral cube was constructed for the spectral window tuned
to the redshifted \cii line.  
The obtained resolution is $0.86'' \times 0.67''$ PA = 94$^\circ$ for A383-5.1
and $1.6'' \times 0.9''$ PA = 84$^\circ$, and in both cases the 
r.m.s. is 11\,$\mu$Jy\,beam$^{-1}$.  
The rms in a 15.6\,MHz channel near the redshifted line is
0.125\,mJy\,beam$^{-1}$ and 0.163\,mJy\,beam$^{-1}$ for A383-5.1 and
MS0451-H, respectively.  

\begin{table}
\caption[]{Summary of the ALMA observations
\label{tab:obssum} }
\begin{tabular}{ccccc}
\hline
\hline
Date & $N_{\rm ant}$ & \multicolumn{3}{c}{Calibrators} \\
 & & Flux & Bandpass & Gain \\
\hline
\multicolumn{5}{l}{\it MS0451-H} \\
10-12-2014 & 37 & J0423-013 & J0423-0120 & J0501-0159 \\ 
26-12-2014 & 40 & Uranus & J0338-4008 & J0501-0159 \\   
26-12-2014 & 40 & Uranus & J0423-013 & J0501-0159 \\   
\multicolumn{5}{l}{\it A383-5.1} \\
31-12-2014 & 36 & J0423$-$013 & J0423$-$0120 & J0239$-$0234 \\
14-01-2015 & 38 & Uranus & J0423$-$013 &  J0239$-$0234 \\
17-01-2015 & 35 & Uranus & J0224+0659 & J0239$-$0234  \\
23-05-2015 & 34 & J0238+166 & J0423$-$013 & J0239$-$0234 \\
23-05-2015 & 34 & J0423$-$013 & J0423$-$013 & J0239$-$0234 \\
24-05-2015 & 34 & J0238+166 & J0224+0659 & J0239$-$0234  \\
24-05-2015 & 34 & J0238+166 & J0423$-$013 & J0239$-$0234 \\
\hline
\end{tabular}
\end{table}

%

\section{Results}

\begin{figure}
\centerline{\includegraphics[width=6.0cm]{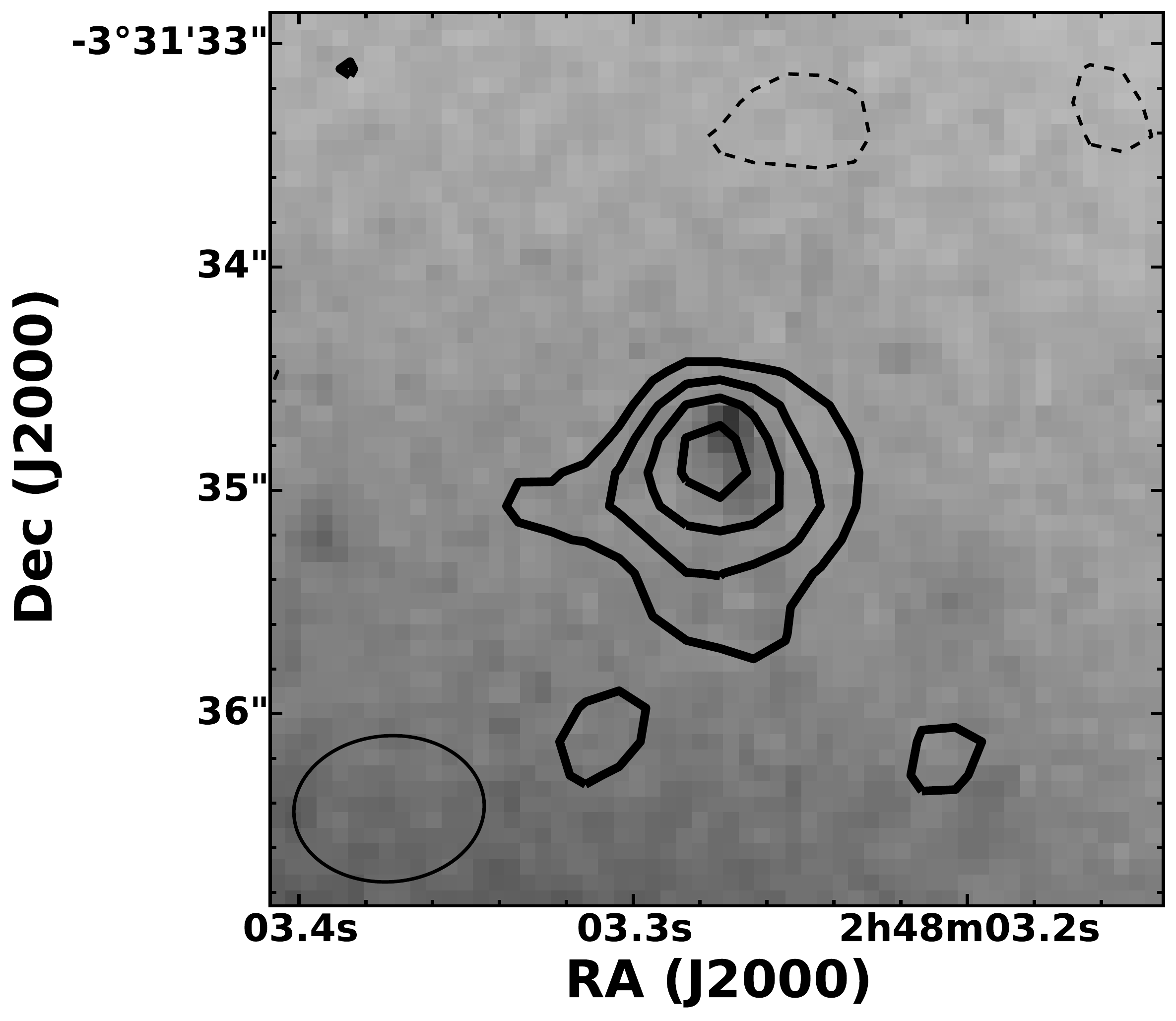}}
\centerline{\includegraphics[width=7.0cm]{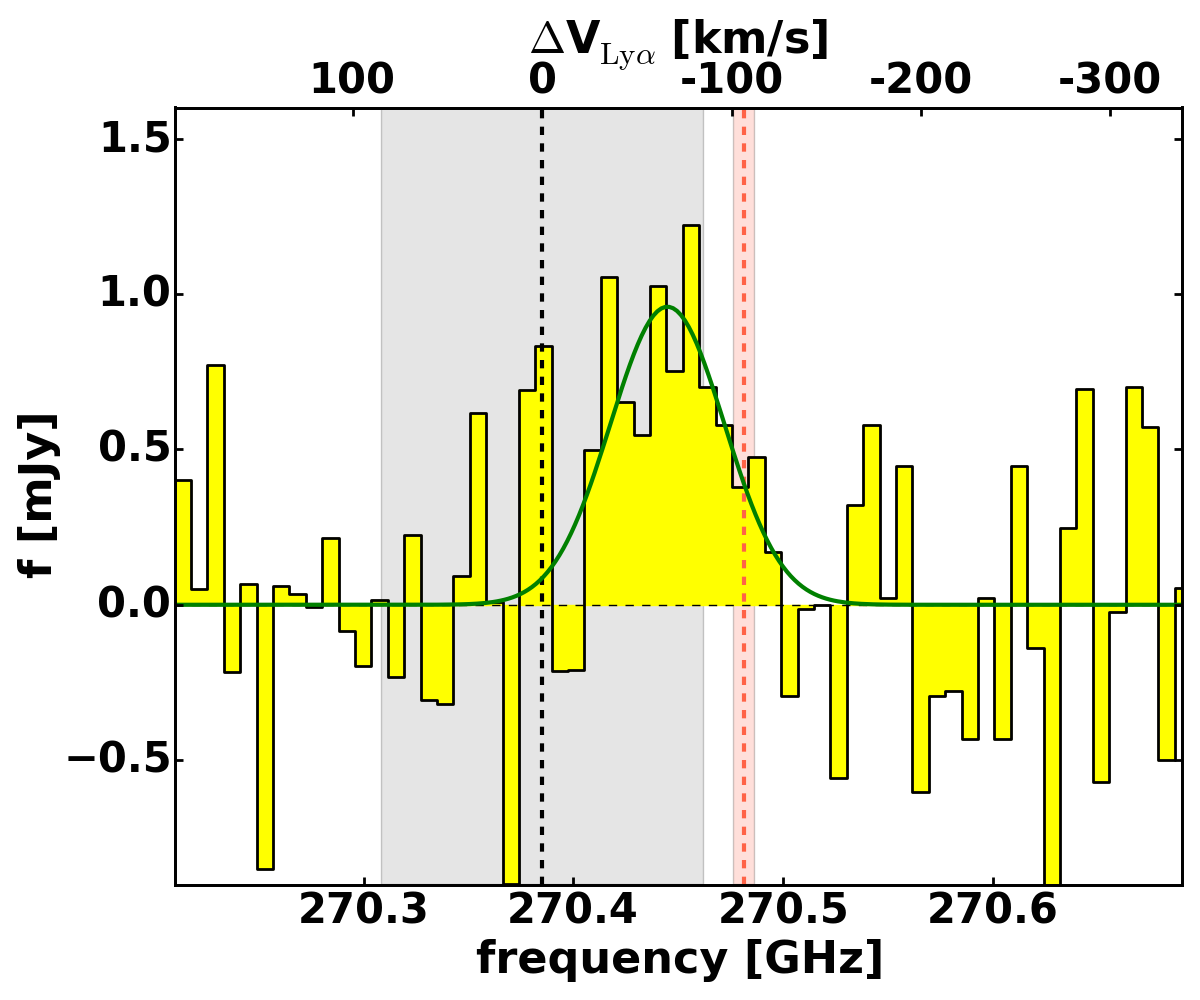}}
\caption[]{
{\it Top}:  HST WFC3 F140W image overlaid with the contours of the
integrated spectral line.  The contours show $2, 3, 4, 5\sigma$, and dashed
show $-2\sigma$.   
The apparent gradient of NIR emission increasing from the lower
part of the image is caused by the bright emission from the central galaxy of
the A383 cluster.  
{\it Bottom}:  
ALMA spectra extracted at the position of A383-5.1 and centered at the
frequency of the redshifted \cii line.  The green solid curve shows the
best-fit Gaussian.  
The vertical dashed line and grey area shows the corresponding redshift and
uncertainty determined from the Ly$\alpha$ line, and the red dashed line and
area corresponds to the redshift measured from C{\sc iii}] \citep{stark15}.
The top-axis shows the velocity relative the Ly$\alpha$ redshift.  
\label{fig:a383result} }
\centerline{\includegraphics[width=7.0cm]{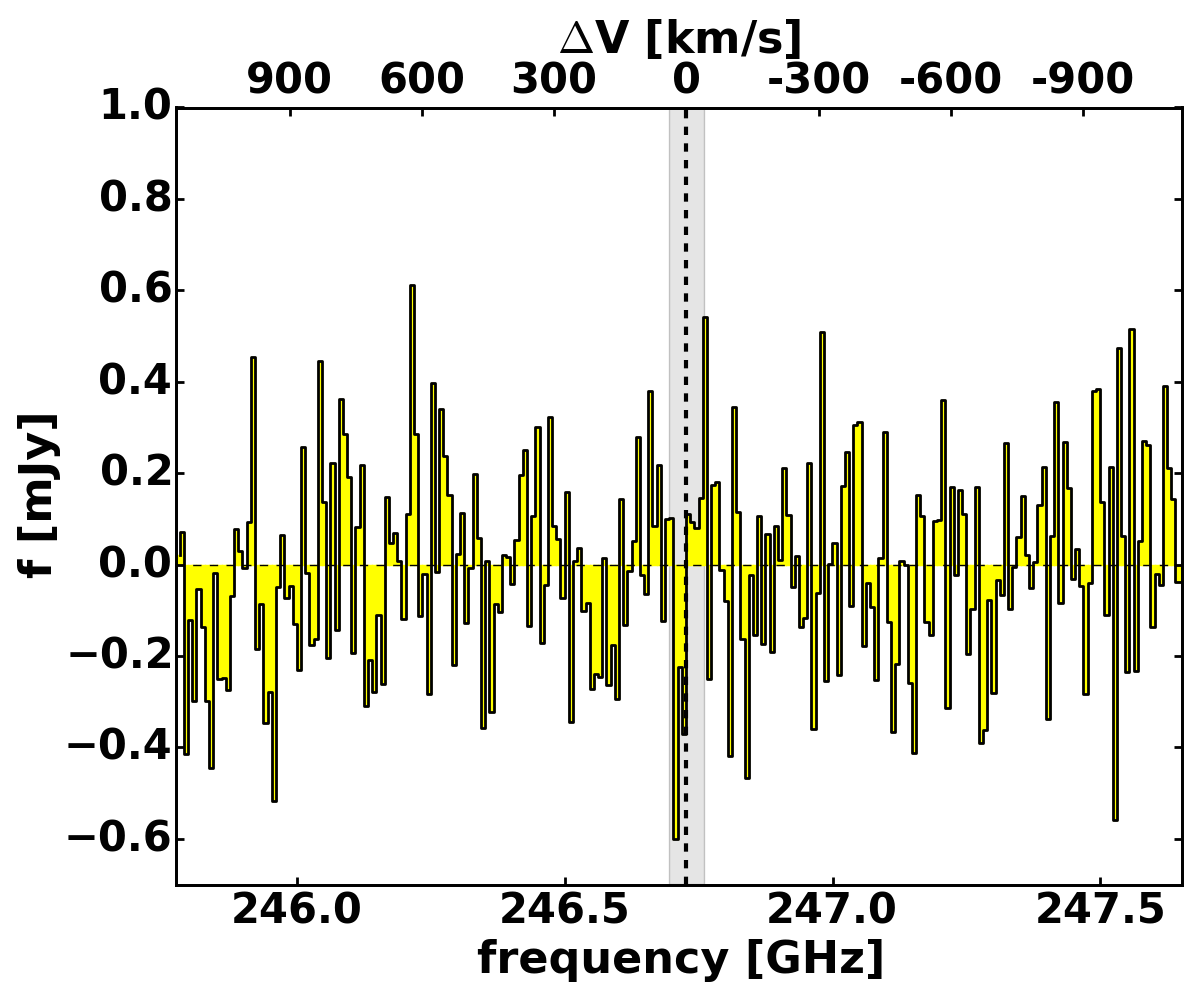}}
\caption[]{The ALMA band-6 spectrum extracted at the position of the MS0451-H
arc.  
The vertical line and grey area shows indicate the corresponding redshift and
uncertainty determined from the Ly$\alpha$ line. The top-axis shows the
velocity relative the Ly$\alpha$ redshift.  
\label{fig:ms0451result} }
\end{figure}

In case of A383-5.1 we detect the redshifted \cii line. The spectrum is
shown in Fig.~\ref{fig:a383result} together with an image of the
integrated line overlaid on a near-infrared {\it HST} F140W image from the CLASH
survey\footnote{https://archive.stsci.edu/prepds/clash/}. 
We fit a
Gaussian line profile to the extracted spectrum and use this to derive the
line properties.  The resulting fit parameters are: central frequency $\nu =  
270.4448\pm0.0061$\,GHz, peak flux of $S_{\rm peak} = 0.96\pm 0.19$\,mJy, and FWHM line width
of $\Delta V = 100\pm 23$\,km\,s$^{-1}$.  This corresponds to an integrated line
intensity $I_{\rm [CII]} = 0.102 \pm 0.032$\,Jy\,km\,s$^{-1}$ (including the
flux uncertainty added in quadrature). 
We derive a redshift of $z_{\rm [CII]} = 6.0274\pm0.0002$, which is in
agreement with the optical/UV redshift of $6.029\pm 0.002$
\citep{richard11} and the {\sc Ciii}] redshift of $6.0265\pm0.00013$
\citep{stark15}.  
We note the spectrum has been extracted from a datacube, which was imaged
using natural weighting and tapering with 2D Gaussian with a corresponding
$1''$ width as the source appears to be marginally extended.  
Using the {\it uv}-data of collapsed spectral channels around the peak of the
emission line, we estimate the size of the source to be $\sim 0.16''$ assuming a
2D Gaussian brightness distribution,
however, due to the modest signal-to-noise ratio it is difficult to
accurately measure its extent.  
The continuum is not detected, and we place a $5\sigma$ upper level
55\,$\mu$Jy.  

In the field around A383-5.1 we also detect two continuum sources, at
02:48:03.4, -03:31:44.86 with a flux of 2.2\,mJy, and at 02:48:02.8,
-03:31:27.67, integrated flux of 0.25\,mJy, and an estimated source size of
$2.4'' \times 1.1''$ PA = 60\,$^\circ$.  The former is associated with the cD
galaxy of the cluster ($z=0.188$), and the latter is
associated with a spiral galaxy at $z = 0.65$ (denoted B18 in
\citealt{smith01}). 

For MS0451-H, we do not detect a line, and 
in Fig.~\ref{fig:ms0451result} we show the extracted spectrum from the
region corresponding to the near-infrared (NIR) source; as for A383-5.1, the
spectrum was extracted from a cube that was imaged using natural weighting
and tapering.  
We measure a rms of 0.05\,mJy\,beam$^{-1}$ for a channel of
100\,km\,s$^{-1}$.  Assuming this corresponds to the FWHM width of the line, we
estimate a $5\sigma$ upper limit of 0.026\,Jy\,km\,s$^{-1}$.
We do not detect the continuum and place an upper
limit of 55\,$\mu$Jy. 

We estimate the \cii line luminosity using $L_{\rm [CII]} = 1.04\times 10^{-3} S
\Delta V D_{\rm L}^2 \nu_{\rm obs}$
\citep[e.g.][]{solomon05,carilliwalter13}.  
\citet{delooze14} has investigated the \cii line as a SFR estimator for local
star-forming galaxies against other probes and for different classes of
galaxies.  We use the relation for low-metallicity galaxies to estimate the SFR
from $L_{\rm [CII]}$ under the assumption that the \cii line traces
star formation in the galaxies. 
Furthermore, we use the continuum upper limit to estimate the FIR 
luminosity upper limits assuming a modified blackbody spectrum with a temperature $T
= 35$\,K and $\beta = 1.6$ \citep[e.g.][]{remyruyer13}.  For comparison with the SFR
derived from the optical/NIR observations, we estimate an SFR from
the $L_{\rm IR}$ assuming a Chabrier IMF \citep[e.g.][]{carilliwalter13}. 
We correct the values for the estimated gravitational
magnification \citep[][Kneib et al., in preparation]{richard11}.   Results
and upper limits are summarised in Table~\ref{tab:luminosity}. 
We note that the CMB temperature is 19.1\,K and 21\,K for $z=6.027$ and
$z=6.703$, respectively.  Following the analysis of \citet{dacunha13}, we
estimate that the CMB heating would increase the temperature by less than a
per cent for the assumed values and that about 14-20\% of the intrinsic
continuum flux density would be missed due to the CMB background emission.

\begin{table*}
\caption[]{Resulting properties for A383-5.1 and MS0451-H together with
UV-optical estimates. 
\label{tab:luminosity}}
\begin{tabular}{lcccccccc}
\hline 
\hline 
Name & $z_{\rm Ly\alpha}$ & $z_{\rm [CII]}$ & $\mu$ & $L_{\rm [CII]}$ &
SFR$_{\rm [CII]}$ & $L_{\rm IR}$ & SFR$_{\rm IR}$  & SFR$_{\rm optical}$ \\ 
   &   &   &  &  [L$_\odot$] & [M$_\odot$\,yr$^{-1}$] & [L$_\odot$] &
[M$_\odot$\,yr$^{-1}$] & [M$_\odot$\,yr$^{-1}$] \\ 
\hline 
A383-5.1 & $6.029\pm0.002$ & $6.0274\pm0.0002$ & $11.4\pm1.9$ & $(8.9\pm3.1)\times10^{6}$ $^a$ & $0.68\pm0.24$ & $<0.5\times10^{10}$ & $<0.5$ & 3.2 \\
MS0451-H & $6.703\pm0.001$ & ...   & $100\pm20$ & $<3.0\times10^{5}$ $^b$ & $<0.04$ & $<0.07\times10^{10}$ & $<0.07$ & 0.4 \\
\hline 
\multicolumn{9}{l}{ $^a$ the error includes the uncertainties from the line
fit, the flux calibration, and the uncertainty on the magnification.}  \\
\multicolumn{9}{l}{ $^b$ assuming a line width of 100\,km\,s$^{-1}$ as
measured for A383-5.1} 
\end{tabular}
\end{table*}

%


\section{Discussion}

\begin{figure}
\centerline{\includegraphics[width=0.90\columnwidth]{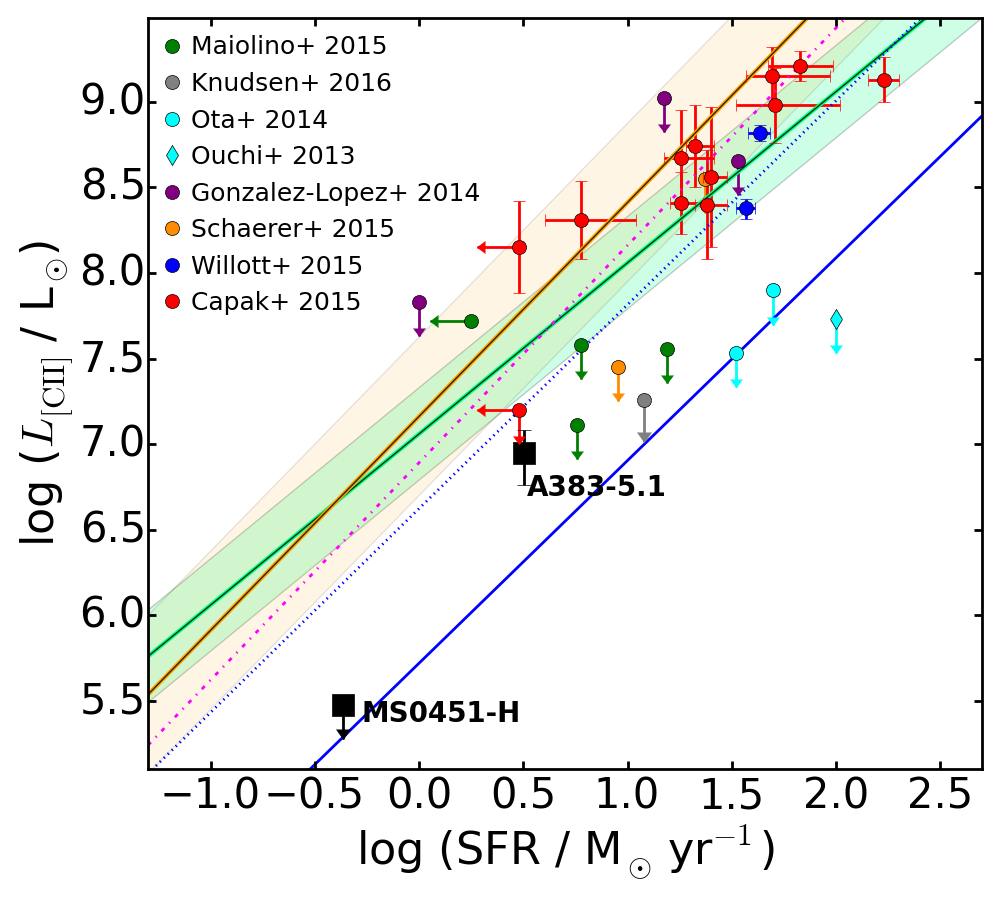}}
\caption[]{
The \cii line luminosity $L_{\rm [CII]}$ vs.\ star formation rate. 
Black squares show the detection A383-5.1 and the sensitive upper limit for
MS0451-H (corrected for magnification). 
We also include the recent $z\sim 6$ results from
\citet{ouchi13,ota14,gonzalez14,willott15b,maiolino15,capak15,schaerer15,knudsen16b}.
The $L_{\rm [CII]}$ - SFR relation, where the green region shows the
relation for local star-forming galaxies and the orange region shows that for
low-metallicity dwarf galaxies \citep{delooze14}.  The blue solid and dotted
line shows the resulting relation from low-metallicity simulations 
\citep{vallini15} (solid: $Z = 0.05 Z_\odot$, dotted: $Z=0.2Z_\odot$) and the
magenta dash-dot-line the results for massive $z\sim2$ galaxies
\citep{olsen15}.   
\label{fig:LciiSFR}}
\end{figure}

A383-5.1 has the lowest \cii luminosity among all detections of this 
line at $z>6$, and the upper limit for MS0451$-$H is more than
1.5\,dex below other upper limits.  Both galaxies are selected as UV-bright,
star-forming, and with gravitational magnification $>10$.  The latter  
enabled us to do deeper observations than previously presented and thus probe
towards SFRs and stellar masses comparable to the low-mass end and less extreme
systems.  In Fig.~\ref{fig:LciiSFR} we show the results together with
results for other high-$z$ and local galaxies. 

Using combined stellar population synthesis modeling and photoionization
modeling of the gas-component, \citet{stark15} finds that the metallicity is
$\log(Z/Z_\odot) = -1.33$ for A383-5.1.  This relatively low metallicity is
comparable to the metallicity found in some nearby dwarf galaxies.  The
the line luminosity for A383-5.1, as well as the MS0451-H upper limit
and the IR luminosity limits are similar to
nearby, low-$Z$ dwarfs'.  
Using {\it Herschel} PACS spectroscopy, \citet{cormier15} studied the
properties of FIR fine-structure lines, including the \cii line, of
the low-$Z$ ISM of dwarf galaxies.   
The $L_{\rm [CII]}/L_{\rm IR}$ ratio has a large scatter for dwarf galaxies,
and the lowest metallicity systems, $Z/Z_\odot < 1/20$, agree with that
scatter.  Based on the $L_{\rm [CII]}$ of A383-5.1, assuming a ratio of 0.6\%
and 0.06\%, we estimate  $L_{\rm IR} \sim
1.4-14\times10^{9}$\,L$_\odot$.  This is partly in agreement with the
lower limit on the $L_{\rm [CII]}-L_{\rm FIR}$ ratio derived from
our results (see Table~\ref{tab:luminosity}), suggesting that the \cii line
contributes significantly to the cooling.  
For MS0451-H, we do not have an estimate for the metallicity given the
limited optical (rest-frame UV) data available (Kneib et al., in
preparation). 

The sample from \citet{capak15} contains bright LBGs with luminosities
$>L^{\ast}$ and in the redshift range $5<z<6$, and while this is below the
redshift range that we are probing ($z>6$), it is to date the largest sample
of \cii detections of galaxies that are not quasar host galaxies nor
SMGs.  That sample is in reasonable agreement with the SFR$-L_{\rm [CII]}$
relations seen for local galaxies \citep[e.g.][]{delooze14}.  Similarly, the
two LBG detections from \citet{willott15b} also follow these relations.  On
the other hand, several \cii searches towards both LAEs and LBGs have
resulted in non-detections \citep{ouchi13,ota14,gonzalez14,maiolino15,schaerer15},
where several observations provide upper limits below the local
SFR$-L_{\rm [CII]}$ comparable to our detection of A383-5.1.  

With an excitation temperature of 91.2\,K and the ionization potential of
carbon of 11.2\,eV, the \cii is a good tracer of the diffuse ISM, of 
the cold neutral medium (CNM) phase as well as of the photon-dominated regions
(PDRs) caused by star-formation in molecular clouds.  
Both models and observations of nearby galaxies show that \cii emission is an
efficient cooling line and that the line luminosity is correlated with the
star-formation rate of galaxies \citep[e.g.][]{delooze14,sargsyan14},
however, for IR bright starburst galaxies and AGN, the efficiency drops
\citep[e.g.][]{diazsantos13}. 

It is unclear why so many $z>6$ galaxies remain undetected, or as in the
case of A383-5.1, show a fainter than predicted \cii line.  
It has been suggested that low metallicity is the main reason.  As shown in
PDR-modelling, the \cii line intensity decreases for lower metallicity,
although not linearly \citep[e.g.][]{rollig06}.  
Given that the \cii line correlates well with the SFR for nearby,
low-metallicity galaxies \citep[e.g.][]{delooze14}, it would be expected that
this is also the case for high-$z$ low-metallicity galaxies.  However, the \cii
deficit suggests that the physical conditions are very different from 
nearby low-$Z$ galaxies.  This \cii deficit is different from the classical
one found for FIR luminous galaxies \citep[e.g.][]{luhman03,stacey10}.  
Aside from metallicity playing a role, other mechanisms could contribute, 
such as a harder radiation field which could
further ionize the carbon into C$^{++}$; in fact the spectral modeling of
A383-5.2 (the counterimage) yields a high ionization parameter log$U = 1.79$
\citep{stark15}.  In such case, observations of
e.g. C{\sc iii}] lines would reveal elevated ratios.  
Increased temperature of the ISM and PDRs would result in other
fine-structure lines being more efficiant coolants and thus brighter than
\cii\!.  
For example, the [O{\sc iii}]\,88\,$\mu$m line is normally expected to be
tracing H{\sc ii} regions rather than diffuse and clumpy inter-cloud medium
\citep[see e.g.][for further details]{cormier15}.  Further, we note that
gas-phase carbon could be depleted onto carbonaceous dust grains. 

In simulations of high-$z$ galaxies, the recent study by \citet{vallini15}
focused on differences arising when taking into account varying metallicity.  
\citet{vallini15} produces a metallicity dependent relation for $L_{\rm
[CII]}$-SFR based on a constant metallicity distribution throughout the
simulated galaxy ($z = 6.6$).  For low $Z$, $Z = 0.05-0.2$\,Z$_\odot$, the
relation is below that of local galaxies.  A383-5.1 has a metallicity similar
to the low-$Z$ assumption of the \citet{vallini15} model.  The model results
are consistent with our detection of A383-5.1, which suggests that
metallicity plays an important role. However, given that we only have one
detection and the remaining being non-detections, we cannot
confirm this. 
We show the Vallini et al. model results in Fig.~\ref{fig:LciiSFR}, and for
comparison, include the resulting relation from multi-phase ISM simulations
of $z\sim 2$ massive galaxies by \citet{olsen15}, which tends to follow the
relations seen for local galaxies.  

It is, however, important to also investigate the selection bias for high-$z$
galaxies.  The fact that Lyman-$\alpha$ emitters, despite high estimates for SFR,
remain undetected in the relatively deep ALMA observations of the \cii line,
could suggest that SFRs are over-estimated and while the Ly-$\alpha$
emission likely traces star formation, a fraction of it
could be powered by other mechanisms, such as shocks of gas inflow.  

To summarize, while it has been suggested that \cii would be a relatively bright tracer
for star formation in high-$z$ galaxies -- and with the advantage of being
redshifted into the $\sim 1$\,mm wavelength range -- the recent results show
that the line luminosity does not follow the same correlations as local
galaxies.  With the A383-5.1 detection, we find that it is possible to
detect \cii towards $z>6$ star-forming galaxies, although the line luminosity
is lower than predicted from local galaxies and also from high-$z$ SMGs and
QSO hosts.  This implies that future observations have the
potential to yield detections, although the predicted line flux will likely
be lower than when derived from local relations, even when using local
low-$Z$ relations. 

\section*{Acknowledgements}
We thank the staff of the Nordic ALMA Regional Center node for their support
and helpful discussions. 
We thank the referee for useful comments. 
KK acknowledges support from the Swedish Research Council (grant:
621-2011-5372) and the Knut and Alice Wallenberg Foundation.  
JR acknowledges support from the ERC starting grant CALENDS (336736). 
JPK acknowledges support from the ERC advanced grant LIDA and from CNRS.
MJ is supported by the Science and Technology Facilities Council [grant
number ST/L00075X/1 \& ST/F001166/1].
  This paper makes use of the following ALMA data:
  ADS/JAO.ALMA\#2013.1.01241.S. ALMA is a partnership of ESO (representing
  its member states), NSF (USA) and NINS (Japan), together with NRC
  (Canada) and NSC and ASIAA (Taiwan) and KASI (Republic of Korea), in 
  cooperation with the Republic of Chile. The Joint ALMA Observatory is 
  operated by ESO, AUI/NRAO and NAOJ.





\bibliographystyle{mnras}
\bibliography{cp_z6}

\bsp	
\label{lastpage}
\end{document}